\documentclass{article}

\usepackage{arxiv}
\usepackage{natbib}

\usepackage{eurosym}
\usepackage[utf8]{inputenc} 
\usepackage[T1]{fontenc}    
\usepackage{hyperref}       
\usepackage{url}            
\usepackage{booktabs}       
\usepackage{amsfonts}       
\usepackage{nicefrac}       
\usepackage{microtype}      
\usepackage{graphicx,wrapfig}
\title{Developing a Cost-Effective Radiometer for Fireball Light Curves}

\author{
  Stuart R. G. Buchan$^1$\thanks{$^1\,$School of Earth and Planetary Sciences, Curtin University, Perth, WA 6845, Australia. $^*$ corresponding author: \texttt{stuart.buchan@curtin.edu.au}. $^2\,$Department of Mechanical Engineering, Curtin University, Perth, WA 6845, Australia}
   \And
 Robert M. Howie$^1$ \\
   \AND
   Jonathan Paxman$^2$ \\
   \And
  Hadrien A. R. Devillepoix$^1$ \\
}
\date{submitted to the proceedings of the International Meteor Conference 2018, Pezinok}

\begin{document}
\maketitle

\begin{abstract}
Fireball light curves can give insight into the meteor ablation process which can be used to improve fireball trajectory and mass modelling. To this aim, the Desert Fireball Network (DFN) is developing a low cost add-on fireball radiometer to supplement existing observatories. The objective is to collect radiometric data on fireballs across a wide spectral range at 1000 samples per second with sensitivity to a large dynamic range (mV  $\in$[-4, -20]), whilst maintaining a low cost. Here we discuss the current prototype design and first light results.
\end{abstract}

\section{Introduction}
	Due to the often very rapid period of ablation in the upper atmosphere, the light intensity of a meteor can be difficult to characterise using solely photographic means. This shows the necessity of a standalone, high-accuracy radiometric device to supplement existing cameras in a fireball network. The implementation of radiometry in the production of meteor light curves provides the unique advantage of a fast sampling rate, unparalleled by photometric equipment, that allows the device to accurately portray rapid changes in brightness. Moreover, through the use of inexpensive silicon photodiodes, the spectral response range can be larger than that of photographic detectors. This means that radiometry possesses sensitivity to wavelengths that could otherwise be neglected by photographic cameras. An example of this can be seen in Figure \ref{specresp}, which demonstrates high sensitivity in the near-IR spectrum. Furthermore, it can be seen that radiometry shows good coverage of the 777 nm oxygen line \cite{jenniskens2018detection}. As radiometers can detect fireball events in cloudy conditions, they have been proven to supplement photographic methods to increase throughput of meteor detection \cite{spurny2001common}. Whilst showing high applicability to meteor detection, it is evident that radiometry is not as popular as photometry in fireball networks around the world. In an effort to amend this, it is proposed that inexpensive off-the-shelf circuit components are to be assembled into a low-cost radiometric device suitable for mass implementation in a fireball network.
	
\section{Design}
	Using the recommendations proposed by Denis Vida in his IMC 2015 paper, design goals for the radiometer were set to have 24-bit resolution, a large dynamic range, and a minimum sampling rate of 1 kHz \cite{vida2015low}.

\subsection{Collection Method}
\begin{figure}[htb]
		\centering
		\includegraphics[width = 0.5\textwidth]{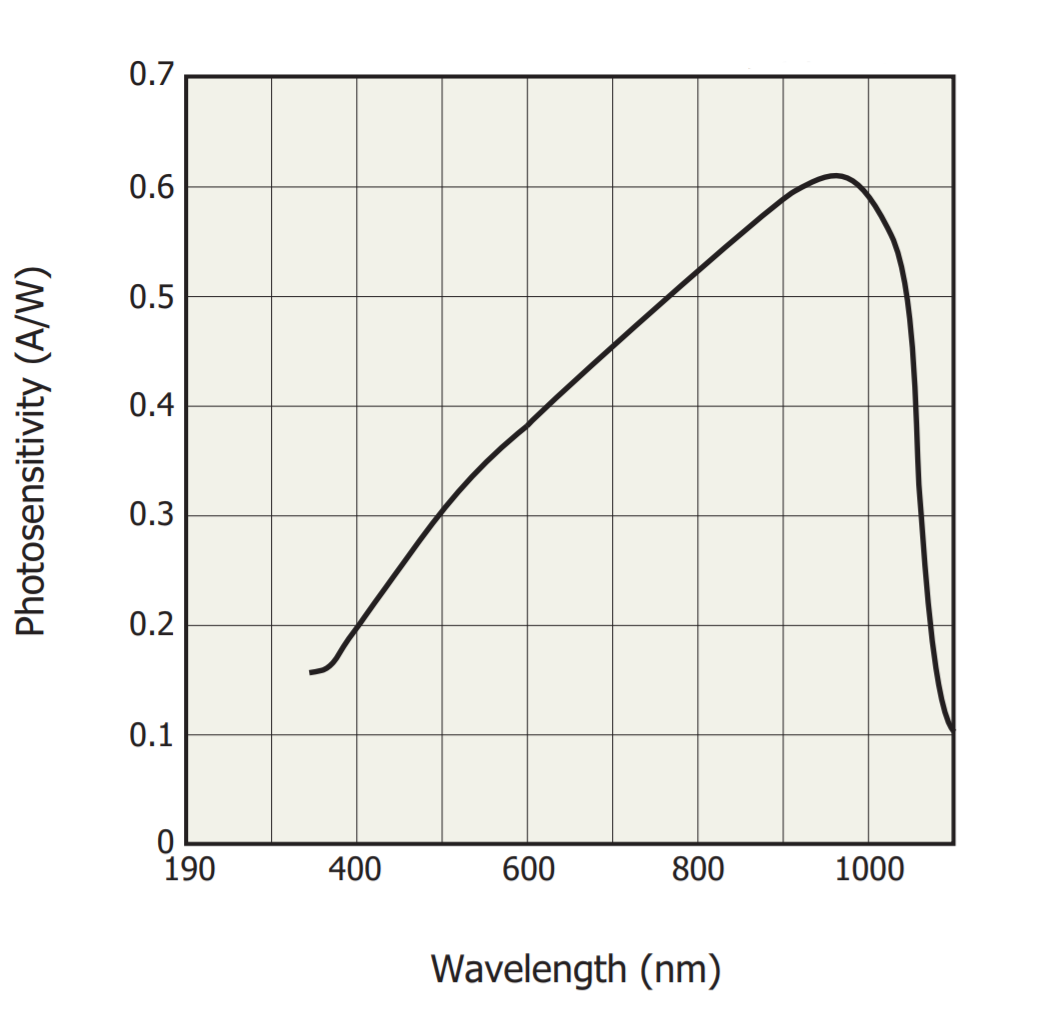}%
		\vspace*{3pt}%
		\caption{Hamamatsu S1337-1010BR Spectral Response\protect\footnotemark}
		\label{specresp}
	\end{figure}

	As an aim for the radiometer is to collect irradiance data over a wide range of wavelengths, silicon photodiodes sensitive in a broad spectral range are desirable. Furthermore, they are seen to possess a very low dark current and high speed response, showing high applicability to the project. At \euro{}50, the Hamamatsu S1337-1010BR silicon photodiode offers an effective photosensitive area of 100 mm$^2$ with a peak photosensitivity of 0.6 Amps/Watt, as can be seen from Figure \ref{specresp}.
	
	\footnotetext{\url{https://www.hamamatsu.com/resources/pdf/ssd/s1337_series_kspd1032e.pdf}}
	
	The irradiance (in $W/m^2$) of a fireball as a function of apparent magnitude can be calculated by the following:
	
	\begin{equation}
		\label{irradiance}
		E = 1100*2.512^{-26.7-M_{fb}}
	\end{equation}
	
\begin{figure}[htb]
		\centering
		\includegraphics[width = 0.5\textwidth]{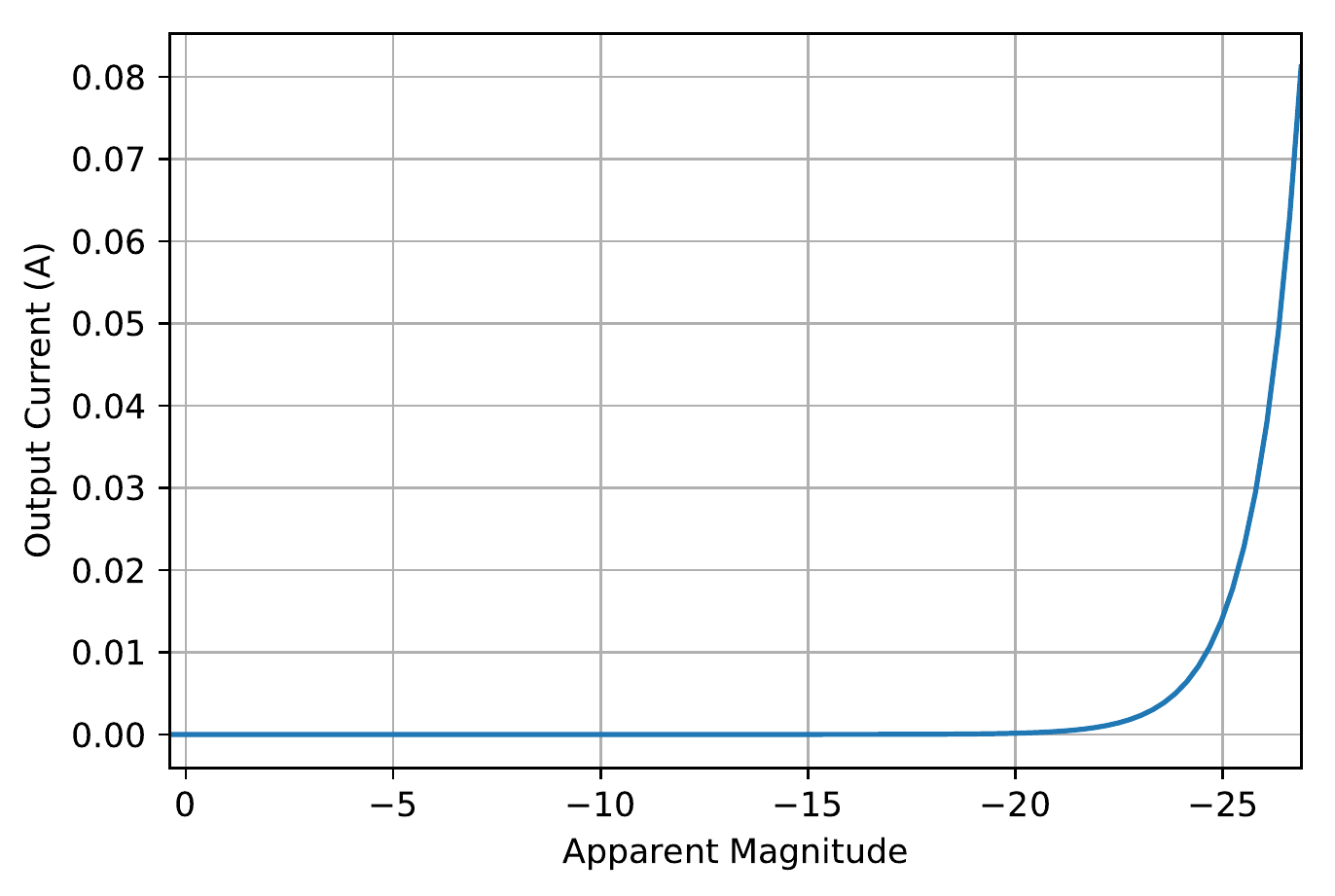}%
		\vspace*{3pt}%
		\caption{Theoretical current output over the range mV $\in$[0.4, -26.9]}
		\label{currentOut}
	\end{figure}

	Where $M_{fb}$ refers to the apparent magnitude of the fireball \cite{spalding2017photoacoustic}.	In an effort to simulate the response of the chosen photodiode to impinging light, a Python script was used to iterate through the dynamic range of interest and calculate the output current using the irradiance derived through Equation \ref{irradiance} with the effective photosensitive area and peak spectral response. The result can be seen in Figure \ref{currentOut}.
	
\subsection{Amplifier}
\begin{figure}[htb]
		\centering
		\includegraphics[width = 0.5\textwidth]{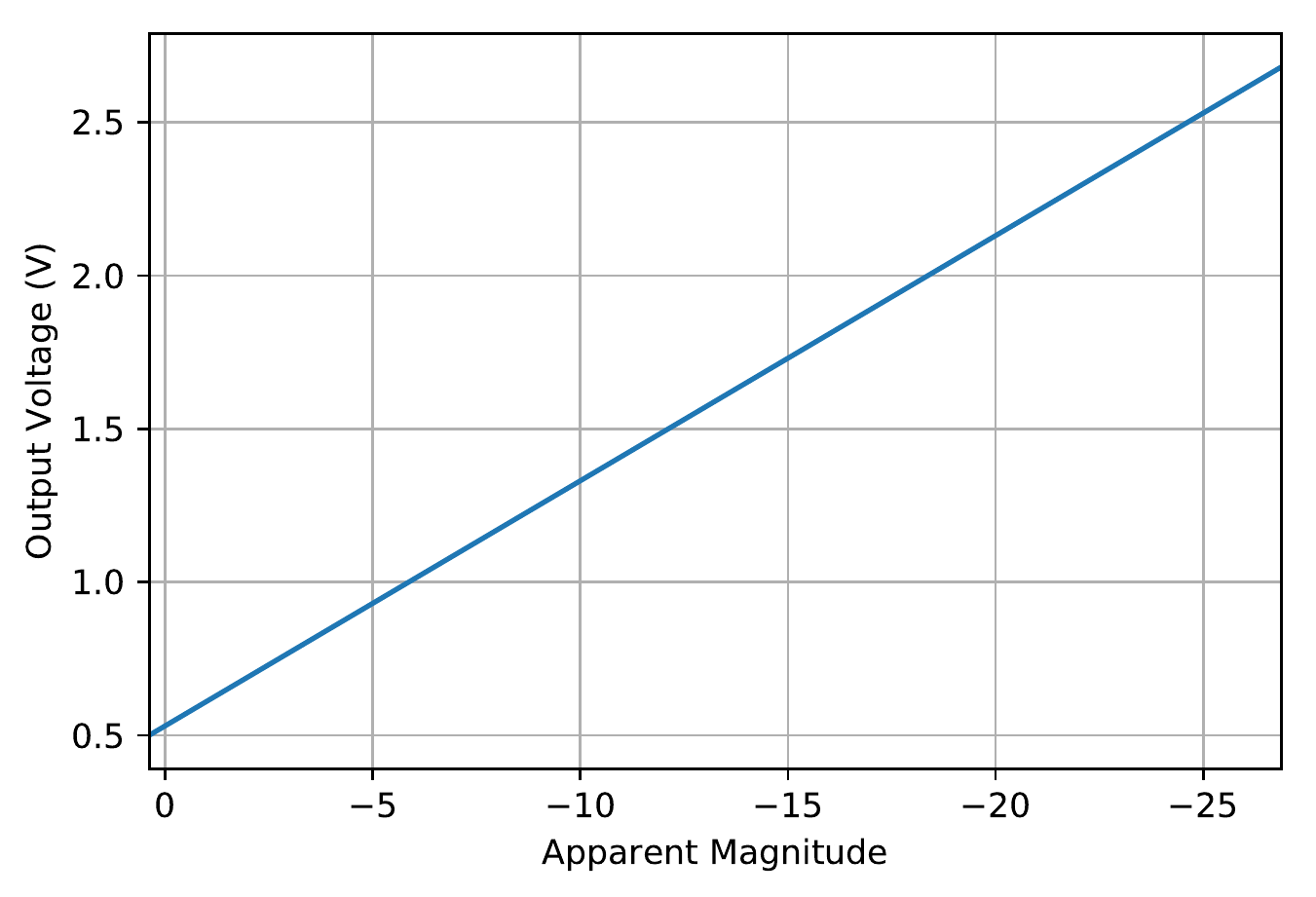}%
		\vspace*{3pt}%
		\caption{Theoretical voltage output over the range mV $\in$[0.4, -26.9]}
		\label{voltageOut}
	\end{figure}

	The purpose of the amplifier is to not only amplify the signal to a recordable level, but to also convert the current signal coming from the photodiode into a voltage that can be sampled. As is evident from Figure \ref{currentOut}, the magnitude of the output signal from the photodiode is insignificantly small for the majority of the dynamic range before ramping suddenly towards the higher end of the range. As a design goal is for a high saturation point, it was decided that a fixed amplifier gain would not be suitable as it would lead to premature circuit saturation before reaching the end of the dynamic range desired. Rather, a logarithmic gain showed strong suitability due to its ability to amplify an input current in a logarithmic manner. The selected component, the Analog Devices ADL5304, offers ten decades of input current from 1 pA to 10 mA, showing strong applicability to the project. The chosen circuit configuration gave a slope of 200 mV per decade, and the output was simulated using the currents seen in Figure \ref{currentOut} to produce the response seen in Figure \ref{voltageOut}.
	
\subsection{Analog to Digital Converter (ADC)}
	To store the voltage signal produced by the amplifier, it needs to be converted into an integer through the use of an analog to digital converter. The chosen device, the Texas Instruments ADS1255, is a low-noise sigma-delta ADC which offers 24-bit resolution and supports sampling rates up to 30 kS/s. For the current prototype design, the ADC has been programmed to sample at 1 kS/s due to limitations of the circuit readout hindering conversion speed. The chosen package outputs the 24 bit sample as three successive bytes via a serial peripheral interface.

\subsection{Readout}
	To remain within the design criteria of low cost and USB accessibility, it was decided that the readout of the radiometer was to be implemented using an inexpensive Arduino Micro microcontroller. The microcontroller communicates with the radiometer through the ADC, and hence needs to conform to the strict timing procedures detailed in the data sheet for the ADS1255. The initialisation and communication protocol was coded using Embedded C. After receiving and converting the two's compliment sample from the ADC, the microcontroller then outputs the radiometric data over USB serial. To log the data, a Python script was written utilising the PySerial library to access the system serial ports. The script when executed waits until the top of the UTC minute, and then starts receiving samples when they are available on the serial buffer. In the current prototype design, the script time-stamps each sample using the system clock, however future revisions will see the implementation of GNSS timing. After storing these values in memory until thirty seconds has elapsed, the script then writes the values to the disk. At the time of the IMC conference, this was done in csv format. However, this has since been updated to logging using FITS tables.
	
\begin{figure*}[!t]
	\centering
	\includegraphics[width = \textwidth]{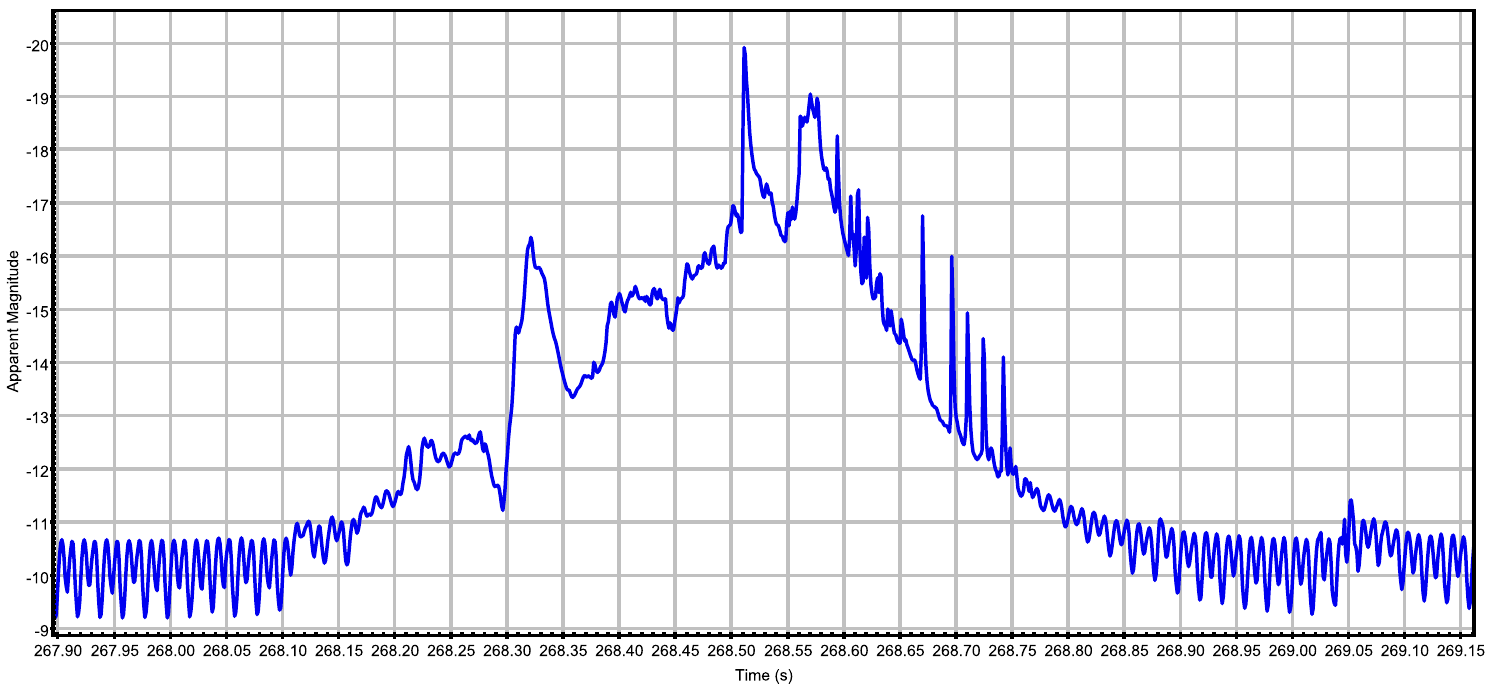}%
	\vspace*{3pt}%
	\caption{Light curve of a lightning strike. Plotted against seconds from starting time of 2018-09-01T20:53:00.0 UTC}
	\label{light1}
\end{figure*}

\section{Implementation}
\begin{figure}[htb]
	\centering
	\includegraphics[width = 0.5\textwidth]{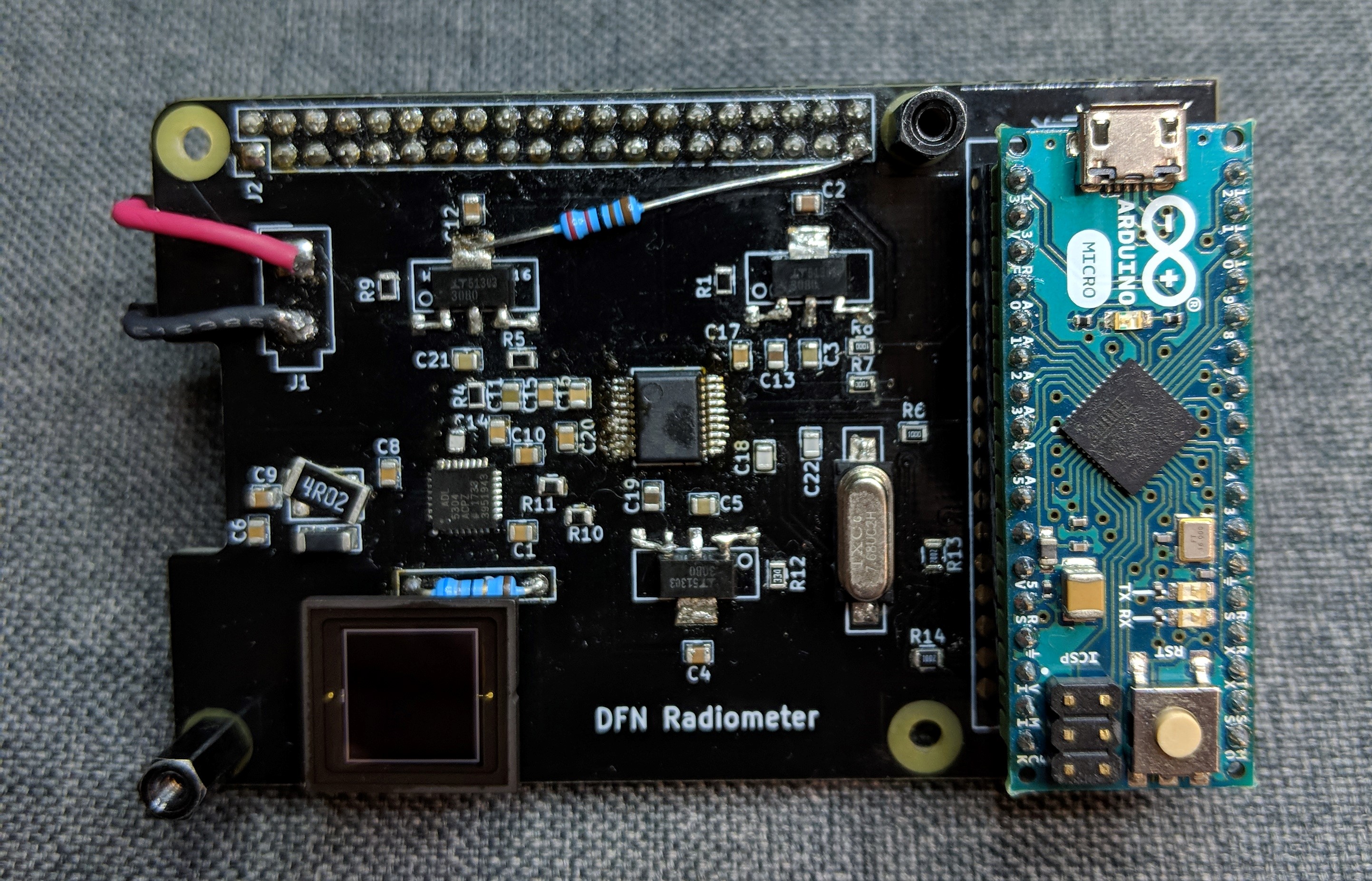}%
	\vspace*{3pt}%
	\caption{Assembled radiometer prototype}
	\label{radiometer}
\end{figure}

	The assembled prototype can be seen in Figure \ref{radiometer}. The completed radiometer is sensitive to a wide spectral range (ultraviolet to near infrared) for under \euro{}100. Due to the large collection area and amplification method, it is able to reliably sample light signals within the range mV $\in$[0.4, -26.9], exceeding the desired initial range. The unit can be powered off a voltage source within the range of 6-36 V, consuming less than 3 W. Future revisions will see the radiometer functioning through a single USB cable for both data logging and power.

\section{Calibration} \label{calibration}
	The unit was calibrated initially by mathematically deriving the relationship between photodiode current and output sample by using the ideal equations given in the data sheet for the circuit components. This however showed an inconsistent error of  around three percent, and hence further calibration was done using celestial bodies. Firstly, the radiometer was exposed to the sun with an incidence angle of 90 degrees. Following this, the radiometer was exposed to the moon with an incidence angle of 90 degrees. The apparent magnitudes recorded for the bodies were -24.5 and -11.2 respectively. These apparent magnitudes were cross referenced with the HORIZONS web interface\footnote{ \url{https://ssd.jpl.nasa.gov/horizons.cgi\#results}}, showing the true values of -26.72 and -11.52 respectively. As the errors were not unanimous, a simple scale was not appropriate, and hence a new model was derived using a linear two point approximation of the sources. This approach is different to photomultiplier tube based radiometers which require external photographic measurements to calibrate for each event, whereas this unit can be absolutely calibrated independently.	

\section{Results}
	On the night of the first of September 2018, the last night of the IMC, Pezinok was struck with a thunderstorm. Taking advantage of this event, the radiometer was placed facing out of a window at the sky to capture light curves of the lightning taking place. An example of the light curves collected can be seen in Figure \ref{light1}. The light curve has been calibrated into apparent magnitude using the model discussed in Section \ref{calibration}. The data is plotted in seconds from the start of the recording session at 20:53:00. The oscillation seen in the data is due to the 50 Hz electrical mains frequency emanating from a streetlamp at an approximate apparent magnitude of -10. It is evident that the peak of the strike occurs just below an apparent magnitude of -20, showing unsaturated recordings over ten magnitudes. As can be seen from Figure \ref{light1}, the event begins at 268.10 seconds, and ends at 269.0 seconds, spanning a total length of 0.9 seconds. Due to this length, it is expected that the light curve shows the illumination from multiple strikes happening in succession. 

\section{Conclusions and Future Work}
	The proposed design criteria for the project were met to produce a working radiometer for under \euro{}100. Using the prototype design, it was possible to produce apparent magnitude calibrated light curves to a satisfactory standard. The next revision of the radiometer will see implementation of GNSS timing for better resolution time stamping. Furthermore, the single photodiode will be replaced by an array with a similar integrated collection area, allowing for an increase in frequency response. Future revisions will also see the addition of narrow-band filters around selected spectral lines.	It is therefore deemed that the initial prototype design was a success, and with some improvements would make a valuable addition to any fireball observatory.

\bibliography{template}{}
\bibliographystyle{abbrvnat}

\end{document}